\documentclass[fleqn,10pt]{wlscirep}
\usepackage[utf8]{inputenc}
\usepackage[T1]{fontenc}
\usepackage[capitalise, noabbrev]{cleveref}
\usepackage{comment}
\usepackage{subcaption}

\title{Origami Single-end Capacitive Sensing for Continuous Shape Estimation of Morphing Structures}

\author[1,*]{Lala Ray}
\author[1]{Daniel Geißler}
\author[1,2]{Bo Zhou}
\author[1,2]{Paul Lukowicz}
\author[3]{Berit Greinke}
\affil[1]{German Research Center for Artificial Intelligence (DFKI), Embedded Intelligence, Kaiserslautern, Germany}
\affil[2]{RPTU Kaiserslautern-Landau, Kaiserslautern, Germany}
\affil[3]{Berlin University of the Arts, Berlin, Germany}

\affil[*]{lala\_shakti\_swarup.ray@dfki.de}

\begin{abstract}
In this work, we propose a novel single-end morphing capacitive sensing method for shape tracking, FxC, by combining Folding origami structures and Capacitive sensing to detect the morphing structural motions using state-of-the-art sensing circuits and deep learning.
It was observed through embedding areas of origami structures with conductive materials as single-end capacitive sensing patches, that the sensor signals change coherently with the motion of the structure.
Different from other origami capacitors where the origami structures are used in adjusting the thickness of the dielectric layer of double-plate capacitors, FxC uses only a single conductive plate per channel, and the origami structure directly changes the geometry of the conductive plate.
We examined the operation principle of morphing single-end capacitors through 3D geometry simulation combined with physics theoretical deduction, which deduced similar behavior as observed in experimentation.
Then a software pipeline was developed to use the sensor signals to reconstruct the dynamic structural geometry through data-driven deep neural network regression of geometric primitives extracted from vision tracking.
We created multiple folding patterns to validate our approach, based on folding patterns including Accordion, Chevron, Sunray and V-Fold patterns with different layouts of capacitive sensors using paper-based and textile-based materials.
Experimentation results show that the geometry primitives predicted from the capacitive signals have a strong correlation with the visual ground truth with R-squared value of up to 95\% and tracking error of 6.5 mm for patches.
The simulation and machine learning constitute two-way information exchange between the sensing signals and structural geometry.
By embedding part of the origami surface with morphing single-end capacitive sensors, FxC presents a unique solution that leverages both the mechanical properties of origami and sensing properties of capacitive sensing.
\end{abstract}

\begin{document}

\flushbottom
\maketitle
%
%
\thispagestyle{empty}

\section*{Introduction}
Folding is a unique structural technique to equip planar materials with 3D spatial morphing properties, especially as deployable origami structures that can be folded or unfolded to change their shape and size.\cite{liyanage2013origami}
Origami structures can be found in many engineering disciplines from aerospace, mechanical, robotics, to architecture and biomedical engineering.\cite{doroftei2014deployable, doroftei2018overview, rus2018design, kshirsagar2023origami}
Research into folding in engineering fields is often driven by the goal of producing lightweight and deployable structures, useful for applications in which the target location is difficult or costly to reach. 
While the status of the folding and deploying processes can be tracked by ad-hoc measurements such as on the external actuators (e.g. motors or pulley systems) or strategically placed contacts and switches in engineering solutions, little research has been carried out on monitoring the folding process through mechanisms embedded in the folding structures. 


Capacitive sensors with planer conductive materials are sensitive to the material's geometry deformation and relative motion.
Conductive fabrics as capacitive sensors  \cite{zhou2023mocapose} have shown potential for recognizing and reconstructing the motion and geometry of the morphing structure from sensor signals.
By leveraging machine learning, even complex relationships between the physical geometry and capacitive signals can be modeled.
Different from widely used stretch sensors often based on e-textile materials, \cite{berglund2014washability, enokibori2014human, stewart2017initial} capacitive sensors do not impose strain on the substrate material and thus do not alter the mechanical property of the substrate on the microscopic scale, which is important for the mechanical functions of origami structures.

\begin{figure*}[ht]
  \centering
  \includegraphics[width=\linewidth]{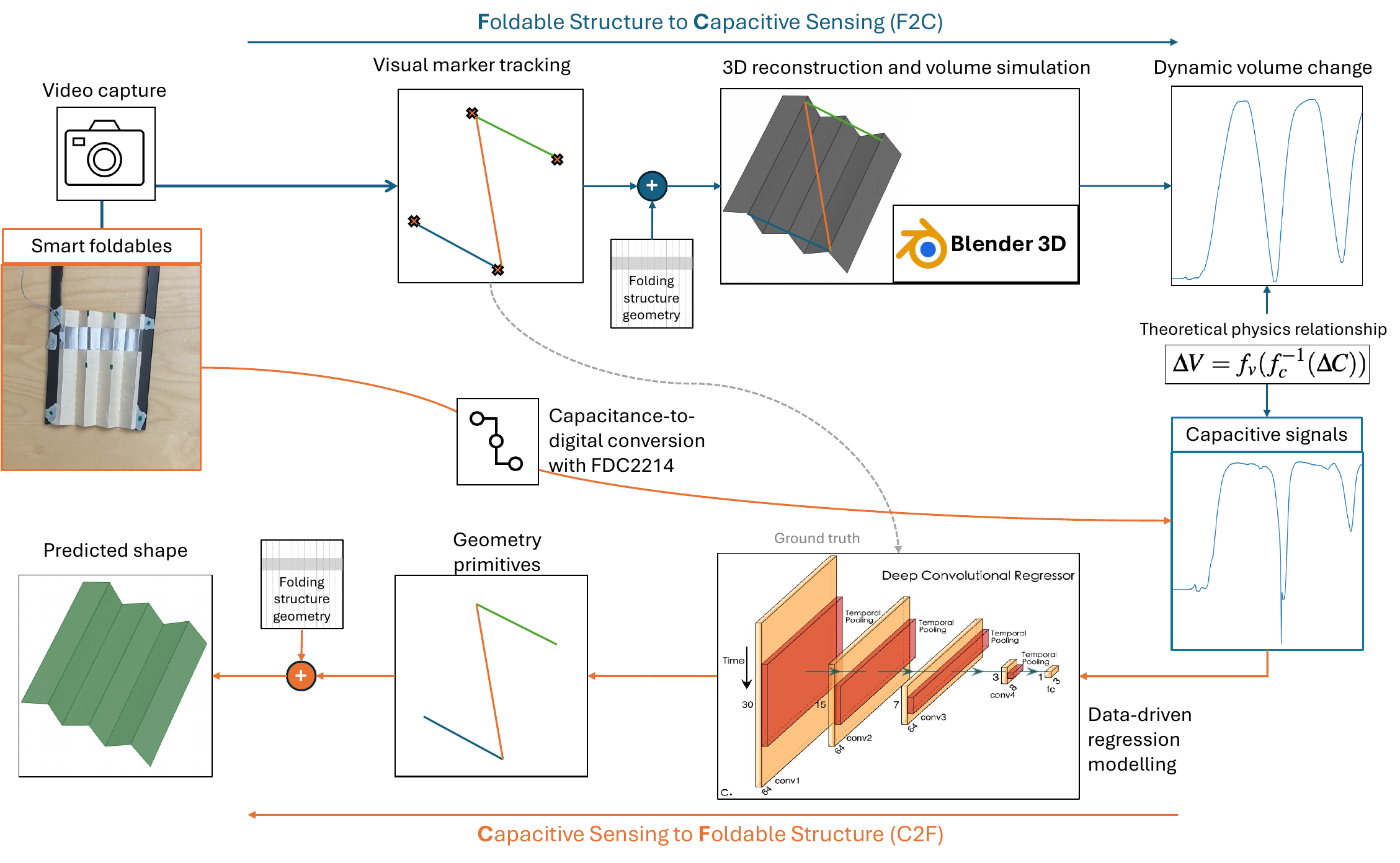}
  \caption{ Overview of FxC that includes the two way pipeline: (a) generating capacitive signal from foldable structure geometry (F2C) and (b) reconstructing the 3D shape of the foldable structure from capacitive signal (C2F).}
  \label{fig:overview}
\end{figure*}

In this work, we propose and evaluate a novel origami sensing paradigm, FxC (Foldable structures with Capacitive sensing), 
using data-driven machine learning to enable morphing structures to track their dynamic folding and deployment process with sensing patches embedded in the folding structures.
As outlined in \cref{fig:overview}, this work can be separated into two coherent parts: the F2C part explains the operating principle of how the origami folding actions can cause predictable single-end capacitive signal changes, i.e. from Folding to Capacitance;
the C2F part implements a shape reconstruction software pipeline from capacitive signals using data-driven deep learning, i.e. from Capacitance to Folding.
We demonstrate that the foldable structures provide repeatable sensor signals with motions.
Additionally through deep learning, the signals can be used to reconstruct the dynamic geometry of the fabric and thus add self-shape-tracking functionality.

The novelty of FxC and the contribution of this paper can be summarized as follows:
\begin{enumerate}
    \item The combination of origami structures with single-end capacitive sensing results in novel shape-changing capacitive sensors that can be used to track the dynamic geometry of the morphing structure.
    \item We tested FxC on different origami structures, both on paper and textile substrates, with conductive tapes and smart textile materials.
    \item Operating principle analysis of the shape-changing capacitive sensors, from folding geometry to capacitance (F2C), through 3D simulation and physics deduction.
    \item From capacitive sensing to track the folding process (C2F), through data-driven machine learning and 3D reconstruction.
\end{enumerate}

\section*{Related Work}
\label{sec:sota}
Morphing structures utilizing folding in engineering are implemented in many domains. For example, in space applications for large-scale foldable devices, such as deployable antennas, \cite{Miura1985b} telescopes, \cite{lightsey_james_2012} solar panels\cite{zhen2017water, Chen2019} or solar sails.\cite{wu2018heliogyro} 
Shape morphing structures are also widely used in biomedical research, \cite{mirzababaei20243d} for example, origami-inspired research has led to small-scale deployable surgical devices utilised for cancer treatment.\cite{Johnson2017} In robotics applications, “4D printing” and origami folded structures allow animation and different modes of motion.\cite{Miyashita2015a, belke2017mori} 
The availability of computational modeling further supports the analysis of dynamic structures \cite{Li2018Foldsketch}, leading also to new methods and applications in design-driven research. \cite{gardiner_ori_2018} 
Digital manufacturing technologies enable foldable structures to be assembled bottom-up, using additive manufacturing methods and smart materials. \cite{Gardiner2018c} 
Folding and pleating have also been investigated in a number of interdisciplinary or design research projects, to study the interaction between folding as arts or scientific practice, \cite{DeRuysser2009a,Friedman2016a} new materials \cite{Philpott2011b} and computational tools. \cite{Gardiner2018c}

Various materials have been studied for folding structures such as laminated composite, \cite{chillara2020review} shape-memory materials, \cite{Hernandez2013Towards, li2021shape} fabrics and textiles. \cite{Knittel2015SelfFolding, sanchez2021textile, xiong2021functional}
One technique to achieve foldable fabrics is pleating, which is done by folding fabric in a regular pattern that is then made permanent by applying heat and steam. 
Different from other textile folding techniques, pleating is applied after the fabric is constructed. It can be done with almost all fabrics and therefore offers extensive versatility regarding textures and volume design.
Mainly implemented in Haute Couture fashion, an example is Miyake’s seminal work on pleating technology where the aesthetic form is paired with textile functionality, e.g. lightness, movement, ease of care, and ease of fabrication. \cite{Miyake2012a}

Textile construction techniques, such as weaving and knitting,  are often chosen for the development of soft circuitry. The bottom-up construction of e-textiles (made directly from yarns or fibres) allows designs that benefit from multi-layered architecture, accommodating controlled interconnection and complex electronic structures. With pleating being a technique that is applied after fabric construction, new soft circuit designs that take advantage of structural properties on fibre, yarn, and fabric levels become possible.  
Despite folding being studied extensively in both textiles and electronics, the combination of the two has been less explored. The majority of examples employ resistive sensors that either require stretching on a structural textile or yarn level, or a complex three-layer design for pressure sensing. \cite{greinke_folded_2022}


It is of importance to estimate the dynamic morphing shape, especially considering control systems.\cite{mcpherson2019dynamics}
Computer vision approach with 3D depth cameras can be applied to track the shape, \cite{minowa2016origami} which requires external sensor setups with sufficient light of sight.
Accelerometers embedded inside an origami structure was demonstrated to estimate the shape of the structure with machine learning. \cite{kinoshita2019shape}

Capacitive sensing using flexible and stretchable materials and microstructures can be used as pressure sensors, strain sensors, proximity, touch sensors, and so on. \cite{qin2021flexible, ma2023recent, cheng2023recent}
The most widely used capacitive sensors are multi-layer parallel-plate capacitors, which leverage the relationship between the capacitance and the distance between the plates as the dielectric layer is deformed by the target physical phenomena and quantities.
Origami structures have been used as the adjustable dielectric spacer layer for parallel-plate capacitive sensors.\cite{liu_wide-range_2021, li2021Paper, sun2023flexible, guo2024bamboo}
Such types of capacitive sensors require three layers: two parallel conductive plates with the middle dielectric layer, which deforms when pressed, resulting in a change of capacity. 
The origami properties behave as springs to return to the original layer thickness after the pressure is leased.
Such capacitors have also been studied for energy harvesting. \cite{gao2018versatile}

Single-layer capacitors can also be used as touch sensors either in a double-plate or single-end capacitor.\cite{salim2017review}
Unlike double-plate capacitors, single-end capacitors use only a single conductor and are also used as proximity sensors to other conductive entities including human body.\cite{ye2020review}
Flexible conductive textiles were also used as capacitive electrodes as patches on a smart wearable jacket `MoCaPose' with 16 channels of single-end capacitors. \cite{zhou2023mocapose}
The patches deform as the wearer moves their body, and the capacitive sensor signals can be used to reconstruct and track the pose of the wearer with machine learning.

To the best of our knowledge, little research has been conducted on utilizing folding materials for shape estimation, especially by integrating capacitive sensing into foldable smart textiles to enable self-tracking capabilities.
\section*{Morphing Capacitive Sensors in Foldable Structures}
In essence, FxC leverages morphing capacitors by embedding conductive patches inside a folding structure.
Thus the capacitor plates follow the folding and extension movements of the foldable structures. 
The conductive patches are connected to a capacitance-digital converter (FDC2214 by Texas Instruments). 
\cref{fig:sim2} shows a minimum setting of FxC in an accordion fold.

\subsection*{Single-end Capacitive Sensors with Morphing Electrodes (F2C)}
\begin{figure}[ht]
  \centering
  \includegraphics[width=\linewidth]{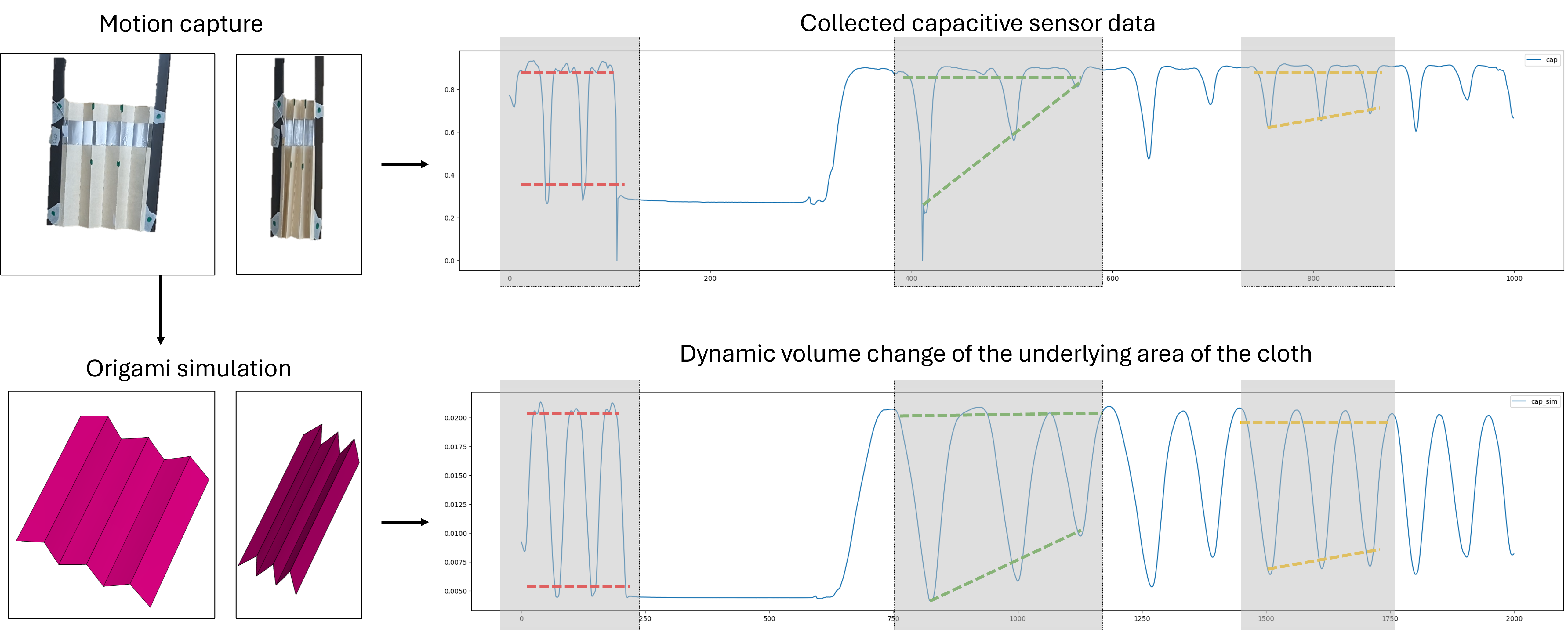}
  \caption{Underlying correlation between captured capacitive signal vs estimated underlying volume of the under the moving Accordion fold (H) with a single channel (Correlation highlighted through the grey-colored overlays)}
  \label{fig:sim2}
\end{figure}

While the physics principle of parallel-plate capacitors is well studied, the principle of single-end capacitors like those in the Theremin are generally treated as antennae, with emphasis on their frequency response instead of the specific capacitance values.   
Also while there exists multiple capacitor simulators, none of the software tools account for single-end capacitors with the plate dynamically morphing its 3D geometry like those embedded in FxC foldable structures.
In this section, we examine the physics principle of such dynamic morphing capacitive sensors, based on knowledge of parallel capacitors.
When the materials are fixed, a parallel capacitor is defined by the geometry of the volume between the parallel plates.
With the assumption that a single-end morphing capacitor should exhibit similar behaviors, we assume a folding structure is placed on a virtual surface plane (representing the virtual ground), and compare the dynamic volume between the folding structure and the virtual plane, with the real captured capacitive sensing data as illustrated in \cref{fig:sim2}.
The simulated folding structure is animated by video captures of the real structure as it is being stretched and folded repeatedly, thus the dynamic volume and capacitive sensor data are synchronized.

The direct output of the FDC2214 capacitance-to-digital converter is a value that is proportional to the measured frequency, which can be converted to the equivalent capacitance by the equation:
\begin{equation}
C = f_\text{f}(\text{f})=\frac{1}{L(2\pi \text{f})^2}-C_{0}
\label{func:ff}
\end{equation}
where $\text{f}$ is the measured frequency, $L$ and $C_0$ are the values of the fixed inductor and capacitor of the analog frontend. 

\begin{figure}[ht]
  \centering
    \includegraphics[width=\textwidth]{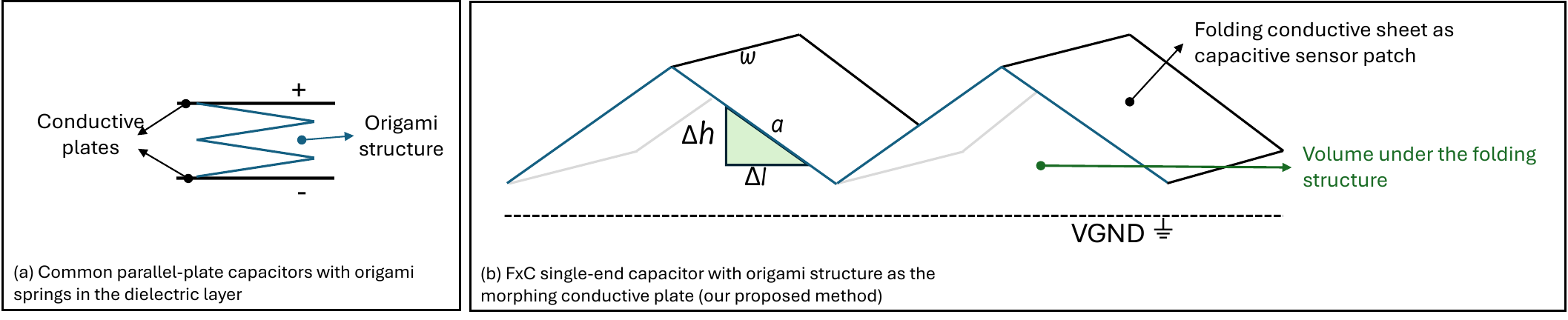} 
  \caption{(a) Comparison with common parallel-plate origami capacitive sensors from the literature. (b) Capacitive sensing principle with single end capacitor and virtual ground. }
  \label{Fig:Cap}
\end{figure}
We can evenly divide the length of the capacitive sensing patch with small constant sections of $a$, then as the structure folds, the cross-section are then divided as small triangles.
Since the small division of the cross section is a rectangle, the height and length are determined by
\begin{equation}
 \Delta l=\sqrt{a^2-\Delta h^2}
 \label{func:l}
\end{equation}
in here since it is a small part of the physical sensor patch, as $l$ and $h$ are variables as the structure folds.

Under ideal conditions, the change of the capacitance of a parallel capacitor can be defined and expressed as a function of $\Delta h$ with the help of \cref{func:l}
\begin{equation}
\Delta C = \frac{\varepsilon A}{\Delta h} = \frac{\varepsilon w \Delta l}{\Delta h} = f_c(\Delta h)=\varepsilon w \sqrt{\frac{a^2}{\Delta h^2}-1}
\label{func:fc}
\end{equation}
where $\varepsilon$ is the electrostatic constant of air, $A$ is the surface area of the electrode, and w is the width of the sensor patch.

Then we can inverse \cref{func:fc} as 

\begin{equation}
\Delta h = f_c^{-1}(\Delta C)=\sqrt{\frac{(\varepsilon w a)^2}{(\varepsilon w)^2+\Delta C^2}}
\label{func:fc1}
\end{equation}

The total volume from the entire patch to the surface underneath which represents the virtual ground plane can be expressed as a function of $\Delta h$ which can be calculated employing 3D simulation Pipeline provided by Ray et. al. \cite{ray2023selecting, ray2024comprehensive}
\begin{equation}
 V =  \sum{\frac{1}{2}w\Delta h\Delta l} 
\end{equation}
and 
\begin{equation}
   \Delta V =  f_v(\Delta h) =\frac{w}{2}\Delta h\sqrt{a^2-\Delta h^2}
   \label{func:fv}
\end{equation}
Combining \cref{func:ff}, \cref{func:fc1} and \cref{func:fv} $\Delta V$ can be written as a function of capacitance change ($\Delta C$):
\begin{equation}
\Delta V = f_v(f_c^{-1}(\Delta C)) = \sqrt{\frac{a^2w}{2} \left (  a^2 \frac{(\varepsilon w)^2}{(\varepsilon w)^2+\Delta C^2} - \left ( \frac{(\varepsilon w)^2}{(\varepsilon w)^2+\Delta C^2} \right )^2\right )
}
\end{equation}
We then replace the groups of constant coefficients to simplify the expression above as 

\begin{equation}
\mathrm{d}V = k_1 \sqrt{ \frac{k_3 k_2}{k_2+\Delta C^2} - \left ( \frac{k_2}{k_2+\Delta C^2} \right )^2 }
\label{func:dV}
\end{equation}

Whereas the constants $k_1 = a\sqrt{w/2} r_1$, $k_2=(\varepsilon w)2 r_2$, $k_3=a^2 r_3$, with $r_1, r_2, r_3$ as the coefficients that alter the ideal conditions in reality.

\begin{figure}[ht]
  \centering
    \includegraphics[width=0.9\textwidth]{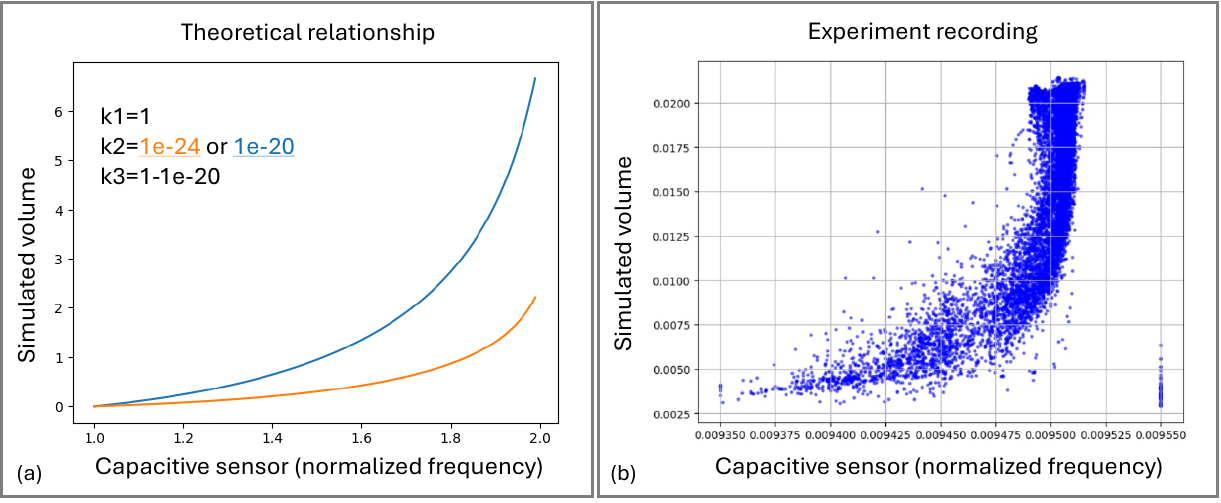} 
  \caption{Comparisons of the principle theoretical relationship and the real collected data of an FXC sample. The constant values were empirically selected.}
  \label{Fig:principle_data}
\end{figure}

When we treat \cref{func:dV} as an indefinite integral of $C$, and combine \cref{func:ff},  the frequency output in the operational range of the circuit (1.0-2.0 $10^7 Hz$), we can plot the curve of the theoretical relationship between the capacitance and volume as in \cref{Fig:principle_data}(a), which exhibits similar behaviors as the experiment recording as shown in \cref{fig:sim2}.
Both show first high capacitance increase with increasing volume at lower ranges, but as the volume further increases, the capacitance approaches saturation.

However, ideal conditions cannot be directly applied to realistic applications as many factors will influence the system such as parasitic capacitance, non-uniform morphing behavior, and even the thickness of the materials.
Thus, in the following, we attempt to predict the geometric changes from the capacitive signals with an ad-hoc data-driven approach.

\subsection*{Data-driven Modelling of the Foldable Geometry from Capacitive Sensing (C2F)}

\begin{figure*}[ht]
  \centering
  \includegraphics[width=\linewidth]{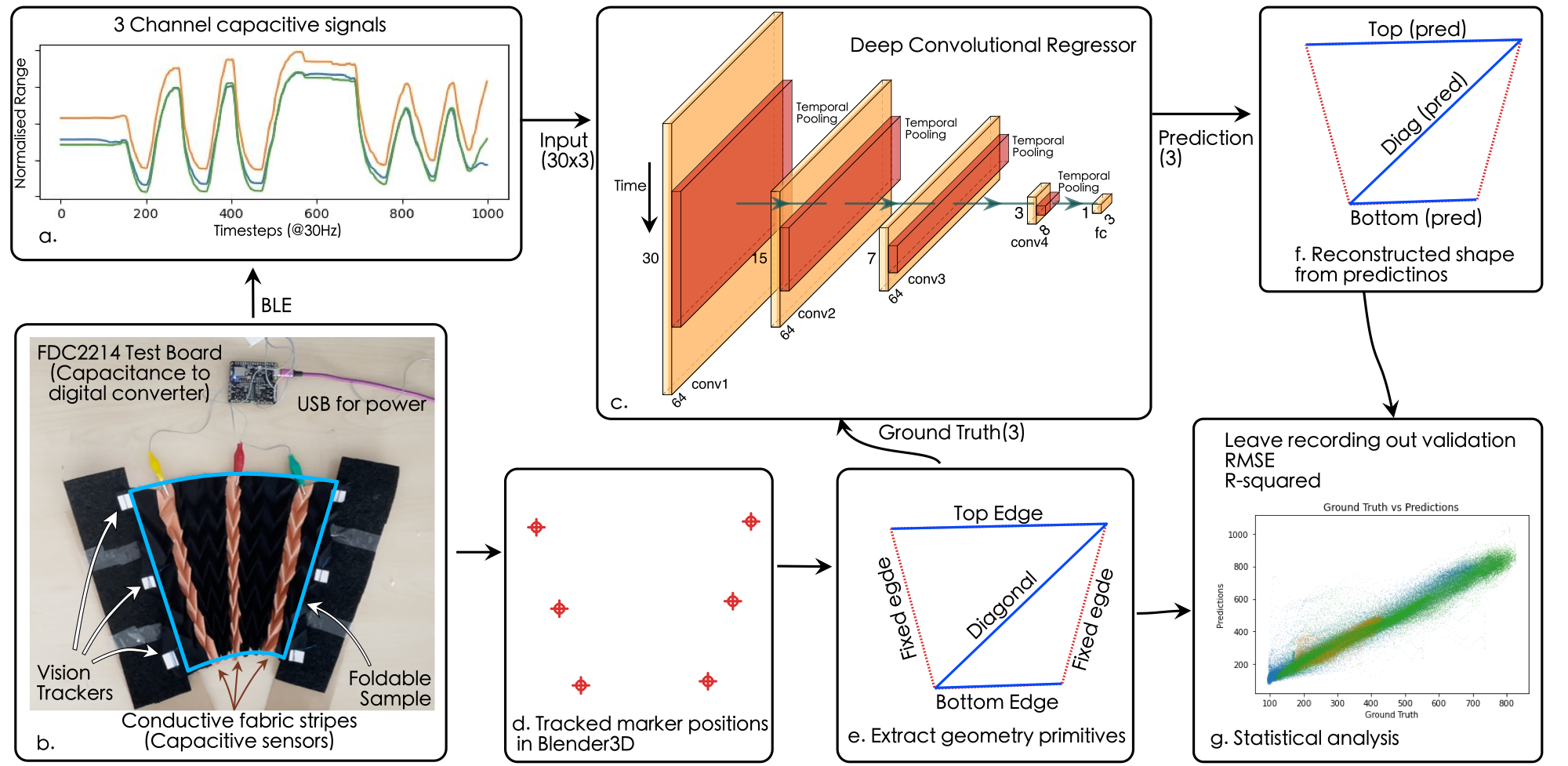}
  \caption{The tracking and DL pipeline for self-shape-tracking with capacitive sensing for Chevron (R) patches.}
  \label{fig:pipeline}
\end{figure*}

While through simulation and physics modeling under ideal conditions, we have observed a clear correlation between the folding structural geometry and the capacitive sensing values, it is not sufficient to apply the principle in broad applications with different folding structures and real-world conditions.
We thus leverage data-driven machine learning to train neural network models to learn the capacitive-to-folding (C2F) relationship, thus predicting the geometry from capacitive signals.
The technical pipeline with an example based on the Accordion (V) patch, one of our utilized sample patterns, is visualized in \cref{fig:pipeline}.

\begin{figure}[ht]
    \centering
    \includegraphics[width=1\linewidth]{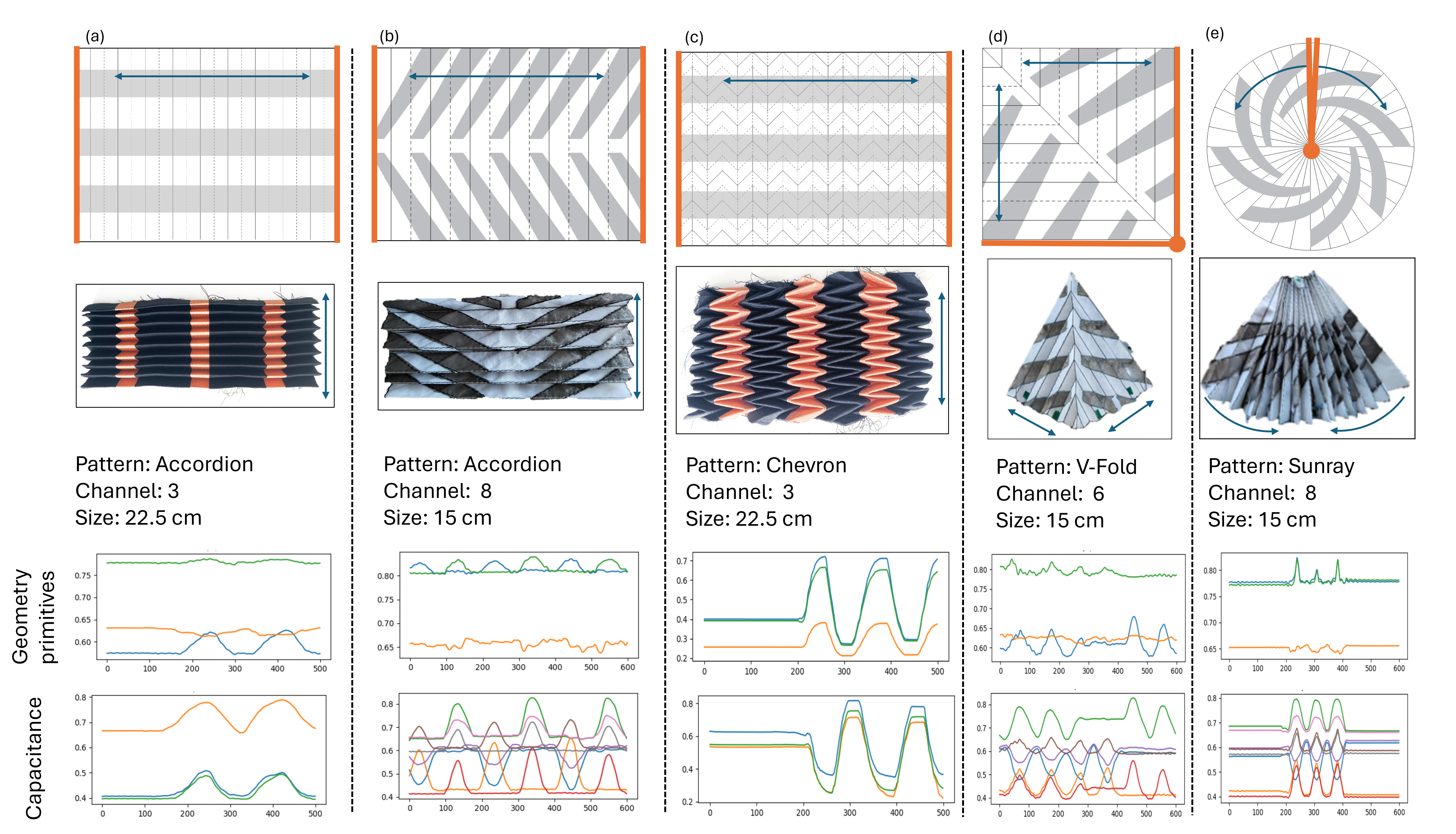}
    \caption{Pleated samples with conductive patches. The arrows indicate the direction of movement, the orange lines indicate the edges fixed to rigid guiding arms. a) Accordion with capacitive patches unfolding perpendicular to folding direction b) Accordion with capacitive patches unfolding diagonally to folding direction c) Chevron pleat with capacitive patches unfolding parallel to folding direction d) V-Fold with conductive stripes unfolding diagonally to folding direction e) Sunray with capacitive patches unfolding diagonally to folding direction.}
    \label{fig:pleated_sample}
\end{figure}

First, the target variables (output) of the ML model need to be defined.
For different origami patterns, we leverage the inherent geometry of the creases as constraints, to simplify the 3D shape to a limited set of geometry primitives, which can be used to reconstruct the overall shape. 
For example, in the case of an Accordion fold, we represent it with three distance values: two representing the two opposite sides of the quadratic foldable structure and one representing the diagonal connecting these opposite sides. 
Similarly, for a V-fold (patterns explained in more details in \cref{fig:pleated_sample}), we utilize three distance values: two representing the connected sides and the third representing the diagonal that forms a triangle of the kite structure. 

\begin{figure}[ht]
  \centering
  \includegraphics[width=0.8\linewidth]{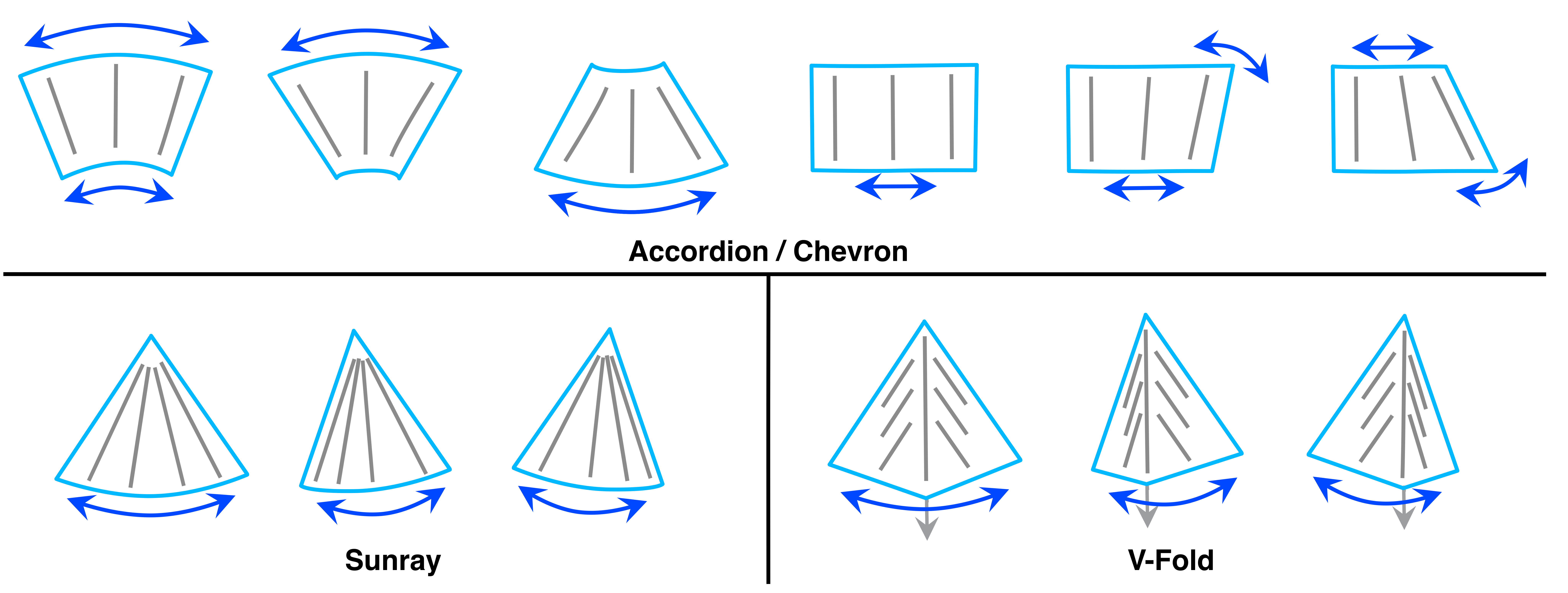}
  \caption{Motion elements during the experiment, the grey lines indicate the directions of the origami ridges.}
  \label{fig:experiment_motions}
\end{figure}

For the training data, experiments are needed to collect synchronized videos of the origami samples and the corresponding capacitive signals from each patch channel, while the origami samples are folded and expanded in different variations as visualized in \cref{fig:experiment_motions}.
Geometry primitives were extracted from each frame of the video with computer vision algorithms.

We then take the capacitive signals as input and the geometry primitives as output, to train a 1D convolution neural network (CNN) regression model.
The trained regression model can be used to predict geometry primitives, which is used to reconstruct the 3D pattern shape with the help of Origami Simulator \cite{ghassaei2018fast} along with Blender 3D. \cite{blender}
The reconstructed 3D model provides a detailed representation of the textile's geometry during various folding scenarios. 



\section*{Evaluation of FxC Origami Structures}
Experiments were conducted to collect synchronized capacitive data and foldable geometry for training and evaluation of the machine-learning model to enable the self-tracking function.

\subsection*{Varieties of Origami Structures and Materials}

We tested four base origami patterns as shown in \cref{fig:pleated_sample}: Accordion, V-fold, Sunray, and Chevron.
Chevron in particular is a repeating V-fold extending in both directions in a zig-zag arrangement. 
A Sunray fold is an accordion pattern with a non-parallel angle, resulting in an arch.
The edges without folding actions of the origami structures were fixed on rigid guiding arms of non-conductive materials. 

The capacitive patches were made from long stripes of conductive materials, placed in such a way the folding action would deform their geometry to form FxC constructs.  
In some samples the capacitive patches were oriented perpendicular or in parallel to the folding direction, we differentiate them with (R) for right angle, or (P) for parallel, for example, Accordion (R) and Accordion (P).
The capacitive patches could also be positioned diagonally to the folding creases, which were implemented with the Accordion, Sunray, and V-Fold.
The analog front end enables multiple channels to operate concurrently without cross-talk.

While the FxC principle of origami single-ended capacitors can be applied to any conductive materials, we validated our approach with commonly accessible materials. While many recent origami and smart textile research involves advanced material science,\cite{de2022use, fang2021smart, libanori2022smart, liu2021smart, luo2021superhydrophobic} the production requirement of such uncommon materials poses a major barrier to technology up-scaling. \cite{zhu2022scaling}
Thus, we propose a two-layered structure using accessible materials and techniques to design the FxC constructs. 
For each origami design we first produced rapid proof-of-concept prototypes with standard paper printed with the pattern as substrate and laminated aluminum tapes as capacitive patches, which were then manually creased to produce the origami pieces. Next, we made a textile version with regular textiles as substrate and conductive e-textiles mounted as capacitive patches.
The e-textiles were fixed on the substrate through either heat-actived adhesive or stitching. 
The origami structure was then created and fixed through pleating.

The FxC constructs were tested by manually moving the arms to expand and fold the samples repeatedly on a flat surface, with a camera positioned perpendicular to the surface to capture the morphing shape of the samples.
The modes of motion of the origami samples are shown in \cref{fig:experiment_motions}.
Visual markers with high color contrast to the rest of the scene were positioned on the guiding arms for shape extrapolation. 

While most samples were tested on a tabletop, the V-Fold samples were suspended on a flat wall.
At idle, the arms hang down due to gravity, folding the sample; while we manually pull the center edge of the V-Fold, the arms will be pushed up by the origami structure, effectively transferring linear movement to angular movement. 

\subsection*{Regression Performance for Shape Reconstruction with Different Patterns}


The Accordion and Chevron samples having rectangular patch shapes are reduced to 3 length primitives \textit{Top}, \textit{Base} and \textit{Diagonal} where first two represents the length of the opposite sides of the rectangular structure and the later is length between their two opposite end of both sides as depicted in \cref{fig:pipeline}. Similarly the V-Fold and Sunray having kite and circular segment shapes are reduced to 3 length primitives where \textit{Left} and \textit{Right} represent two adjacent sides of the triangle formed by them while the \textit{Diagonal} represent the third side of the triangle as depicted in \cref{fig:correlationBA1}. 

\begin{figure}[ht]
  \centering
  \includegraphics[width=\linewidth]{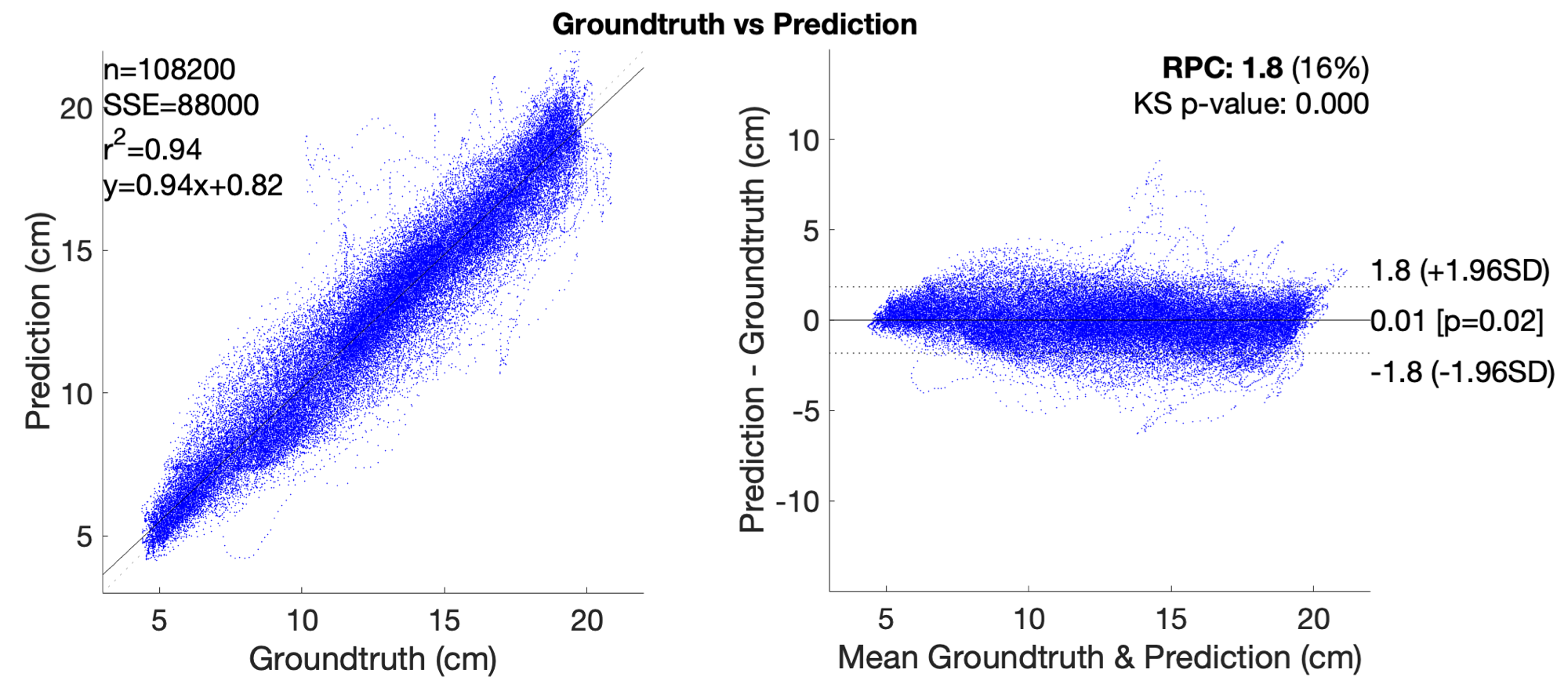}
  \caption{Correlation and Bland-Altman Plot of Accordion(P)}
  \label{fig:correlationBA}
\end{figure}

We refer to two standard statistic metrics to evaluate the regression performance: R-squared ($R^2$) and root mean squared error ($RMSE$).
The regression results and reconstruction errors are detailed in \cref{tab:r2_table}, while the correlation and Bland-Altaman analysis \cite{giavarina2015understanding} is shown in \cref{fig:correlationBA}.
$R^2$ is the coefficient of determination in statistics which describes the goodness of fit of a linear regression between two variables. \cite{kasuya2019use}
$R^2$ of 1 indicates two variables are identical.
In the literature, \cite{boddy2019exploring, auepanwiriyakul2020accuracy} the consensus range of strong correlation is $R^2 \in [0.7, 1]$.
From \cref{tab:r2_table}, the $R^2$ of all variables from all 4  quadratic patches (Accordion, Chevron) are well above 0.85, indicating strong correlation which means the predicted results of the convolutional regressor can be used to reconstruct the quadrilateral outline of the origami samples.
While $RMSE$ measures the error of reconstructed lengths, the average values of are not necessarily smaller with larger corresponding $R^2$ due to outliers, distribution and de-normalisation.
The $RMSE$ of all samples are around 1cm, which is <5\% of the 22.5cm length of the fixed edges of the samples.
The V-Fold patch have $R^2$ of 0.88 and $RMSE$ of 0.65 cm which is around 6.5\% of the reconstructed patch having 10 cm length.
In case of the Sunray it performed worse having an average $R^2$ of 0.35 and $RMSE$ of 5.89 cm, which is >30\% of the length of reconstructed patch.
All sample combinations with both Chevron and Accordion folding patterns, horizontal and vertical conductive stripe layouts, performed similar in the range of 0.4cm, with the Chevron(R) sample slightly outperforming the others.


The Accordion structure with diagonal elements was excluded from the evaluation because of frequent short-circuit between channels and noisy signal.

\begin{table}[ht]
\caption{R-squared for different patterns and RMSE for regenerated pattern shapes. ( $R^2\uparrow$ / $RMSE\downarrow$ in cm )}
\label{tab:r2_table}
\centering
\begin{tabular}{lcccc|lcc}
\toprule
Shape Primitives & Chevron(R) & Accordion(R) &  Chevron(P) & Accordion(P) & Shape Primitives  & V-Fold & Sunray\\
\midrule
\multicolumn{8}{l}{$R^2 \uparrow$}                                 \\
\hline
Top       & 0.95120   & 0.95318     & 0.89667     & 0.96149 & Left &  0.87436 &  0.39875  \\
Diagonal  & 0.82558   & 0.88117     & 0.88533     & 0.95186    & Right & 0.87742 &  0.37125 \\
Base      & 0.90795   & 0.86812     & 0.89229     & 0.94997    & Diagonal &  0.88545 &  0.31507 \\
Average   & 0.89491   & 0.90083     & 0.89143     & 0.95444   & Average & 0.88636 &  0.35250  \\
\midrule
\multicolumn{5}{l}{$RMSE \downarrow$ in cm }                       \\
\hline
Top       & 0.94862    & 1.05803      &  1.35958    & 1.30047  & Left & 0.63910 &  6.78901 \\
Diagonal  & 0.81143    & 1.38266      &  1.10680    & 1.12883   & Right & 0.64183 & 5.36782 \\
Base      & 1.19115    & 1.68110      &  1.57289    & 1.16975   & Diagonal & 0.65521 & 6.22179 \\
Average   & 0.98373    & 1.37393      &  1.34642    & 1.19968  & Average & 0.65499 & 5.89523  \\
\bottomrule
\end{tabular}
\end{table}


\subsection*{Influence of Materials}
\begin{figure}[!ht]
  \centering
  \includegraphics[width=\linewidth]{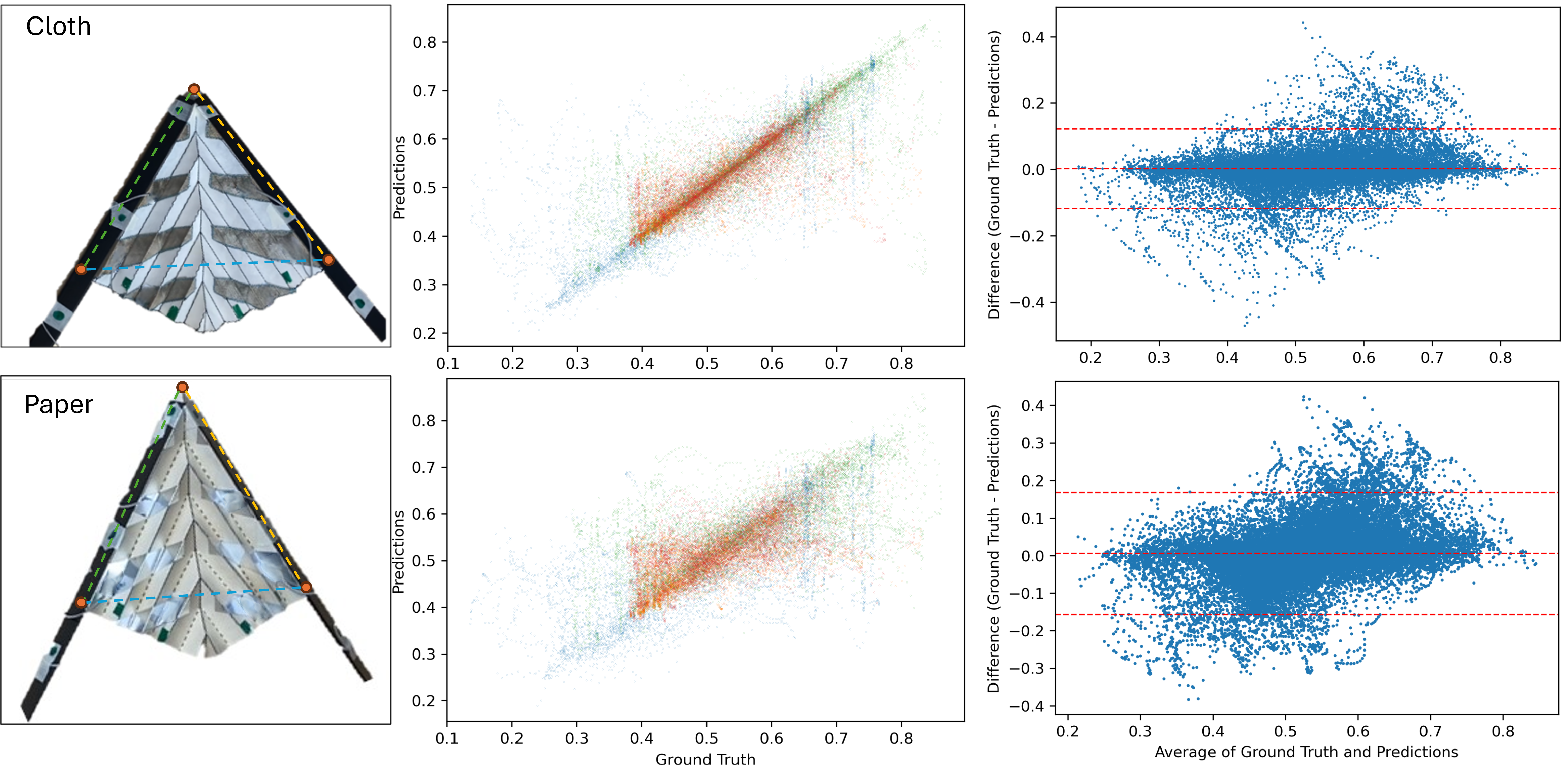}
  \caption{Correlation and Bland-Altman Plot of identical V-Fold patch for different materials.}
  \label{fig:correlationBA1}
\end{figure}

To validate our approach, identical origami structures based on the V-Fold pattern with identical sizes and capacitive strips were tested using the same movement patterns. 
The comparison between cloth origami and traditional paper origami revealed the superior performance of the textile-based system as shown in \cref{tab:r2_table1}. 
The utilized regression model outperforms the prediction on the cloth dataset with almost 40\% $R^2$ at 88\% whereas the RMSE was 0.65cm compared to 0.93 for the paper V-Fold pattern.
Further, we visualized the deviations between paper and cloth through correlation and Bland-Altman plots in \cref{fig:correlationBA1} to emphasize the deterministic relationship between movement and capacitive signal, especially for the cloth sample.

Other than having comparatively better accuracy, textile-based FxC structures also show better shape retention properties, while paper-based FxC structures are more prone to lose the retention of the creases when undergoing folding and unfolding to extreme ranges. 
Although it does not affect the actual capacitive values, it might create other mechanical problems since shape retention degradation may lead to a phenomenon where a different amount of force is required to bring the structure to the exact same shape after different iterations. 

\begin{table}[!ht]
\caption{R-squared and RMSE for same pattern V-Fold made with cloth vs paper with identical size and capacitive strip placement. ( $R^2\uparrow$ / $RMSE\downarrow$ in cm )}
\label{tab:r2_table1}
\centering
\begin{tabular}{lcccc}
\toprule
Variables & Cloth & Paper \\
\midrule
\multicolumn{2}{l}{$R^2 \uparrow$ }   \\
\hline
Left & 0.87436 & 0.47445 \\
Right & 0.87742 & 0.47351 \\
Diagonal & 0.88545 & 0.48538 \\
Average    &\textbf{0.88636} & 0.49445 \\
\midrule
\multicolumn{2}{l}{$RMSE \downarrow$ in cm }   \\
\hline
Left & 0.63910    & 0.92541 \\
Right & 0.64183    & 0.92899   \\
Diagonal & 0.65521    & 0.93297  \\
Average   &\textbf{0.65499}& 0.93154  \\
\bottomrule
\end{tabular}
\end{table}

\subsection*{Limitations and Future Opportunities}
The geometry primitives were used as ML model predictors, with the prior assumptions that the origami structures would follow the designed folding behavior, that the creases remain rigid lines and that the faces surrounded by creases remain flat surfaces.
However, they are only applicable to be used to reconstruct the full 3D shape if the origami patterns are being folded under normal conditions. 
Exceptions are for example being warped outside the plane of designed folding behavior, or damages to the overall structure.

Since the capacitive patches are not isolated on the surface, during the origami folding actions some channels will touch each other and create short circuits.
This problem particularly renders the samples of Accordion fold with diagonal channels, and Sunray pattern extremely unreliable. 
In the work of MoCaPose in wearable capacitive sensing, \cite{zhou2023mocapose} it was discovered that laminating the conductive patches with an isolation layer of thin foil would largely improve the stability.
However, the addition of an extra layer in dedicated origami structures could be more significant than in wearables, which might change the desired folding properties.
It would be interesting to investigate origami single-end capacitors with electrical isolation on both sides.

The placement and number of channels within each pattern currently lack a systematic algorithm or logic. 
Instead, they are determined based on the reasoning behind the movement of the FxC structure, aiming to maximize deformation and capture movement effectively. 
However, implementing a more defined process for selecting channel placement and quantity could significantly enhance the overall performance of the self-tracking property. 
By establishing clear guidelines or algorithms, developers can ensure that channel placement is optimized for capturing movement accurately while minimizing interference and short circuits.
Since we have demonstrated the folding-to-capacitance relationship by simulating the dynamic volume of the origami, the simulation could provide a tool to develop layout optimization methods.

In this work, the FxC constructs were in the range of 10 to 22.5 centimeters in length, which already are significantly larger than other parallel capacitors involving origami structures that focus on micro structures. \cite{liu_wide-range_2021, li2021Paper, sun2023flexible, guo2024bamboo}
While the basic capacitance can be defined by surface area and we show a clear relationship between the origami volume and the capacitance, further experimentation is needed for larger or smaller scales of the FxC method to validate the self-tracking property for specific applications.
Nonetheless, for larger structures for larger structures folds of similar size to the one we have tested can be applied as repeat patterns.

\section*{Conclusion}
Through FxC, we have demonstrated a novel origami single-end capacitive sensing approach with the capacitor plate itself being folded.
In contrast to other origami capacitor studies where the origami was leveraged to construct the spacer dielectric layers of parallel plate capacitors, FxC leverages the shape change of the capacitor plate caused by the dynamic origami folding actions.
The multiple channels of FxC capacitive sensors can be used to track the dynamic shape of the origami structure through deep learning regression to geometry primitives and reconstruction with prior knowledge of the origami layout.
Opposed to other parallel-plate capacitors which are mostly used as pressure sensors, FxC combining multichannel capacitors and regression neural network can be used a shape sensor for the substrate structure.
Several samples with different origami structures based on paper, metal foils, regular textiles and e-textiles were produced.
Through various experiments with synchronized video tracking and capacitive data, our evaluation shows a consistently strong correlation between the capacitive signals and the shape-defining geometry primitives with an average $R^2$ between 0.88 and 0.95 for most of the samples.
The tracking errors after shape reconstruction are around 1cm with samples of 22.5 cm in length and 0.65 cm of the smaller samples of 10 cm in length.
Overall integrating capacitive sensing in origami structures presents a promising solution for dynamic shape estimation with elements embedded in the origami materials.

While precisely tracking the geometry from capacitive patches may become challenging with more complex structures and movements, the capacitive sensing principle \cite{bello2022move, zhou2023mocapose} suggests that the repeatable signal may be exploited for activity classification with pattern recognition machine learning techniques, for example classifying the types of motion through temporal signal characteristics.
FxC structures may also be exploited for structural materials that are aware of the coverage or structural motion in domains such as architecture, robotics and aerospace engineering.
Such self-tracking textile-based smart foldables could also find useful applications in specialized wearable garments with functional folds.

\section*{Methods}
In the following sections, we specify the hardware requirements as well as the implementation procedure in greater detail in order to reproduce our experiments and results.

\subsection*{Folding Smart Textile with Embedded Capacitive Sensor Patches}

For the cloth-based FxC structures, the base fabric of thickness 0.325 mm (Accordion (P,R),  Chevron (P,R)) and 0.1 mm (V-Fold, Accordion (D), Sunray) is made of plain-weave cotton mixed with synthetic fibers. 
Conductive fabric Shieldex\textsuperscript{\textregistered} Kassel RS (Accordion (P,R),  Chevron (P,R) with a thickness of 0.05mm and Shieldex\textsuperscript{\textregistered} Porto RS (V-Fold, Accordion (D), Sunray) having the same thickness, was first fused with an iron-on adhesive (Vliesofix\textsuperscript{\textregistered} Bondaweb\textsuperscript{\textregistered} 719), then laser-cut (alternatively cut manually with scissors or a fabric cutter) into shape, and subsequently thermobonded onto the base fabric.

\begin{figure}[ht]
    \centering
    
    \includegraphics[width=1\linewidth]{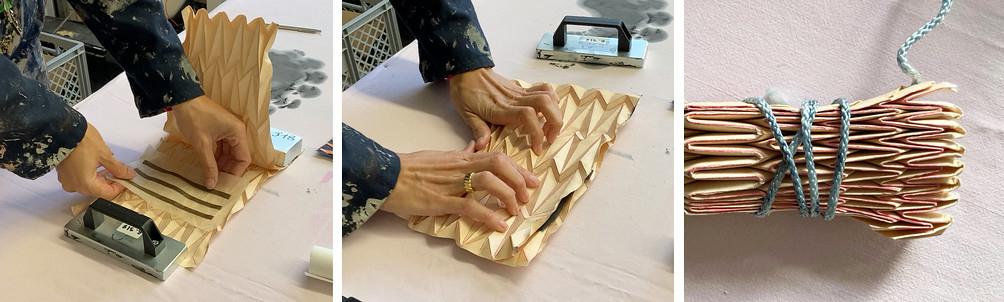}
    \caption{Fabric first layered in between two identical paper molds, then folded together as a sandwich and secured with string.}
    \label{fig:pleating}
\end{figure}

To pleat the fabrics, they were sandwiched between two layers of pre-folded molds made from 120g "Elephant Hide" bookbinding paper. They were then tightly folded and secured with string. The process is shown in Figure~\ref{fig:pleating}. To fix the folds, they were then steamed (UHLIG vertical professional steamer) for 20 minutes at about 120\textdegree C.

Paper-based FxC structures have a cardboard basis with thickness 0.05 mm as a substrate which got attached with TESA 60632 Aluminium stripes, having a thickness 0.075 mm.
Instead of pleating the substrates, we manually folded them into the desired structures.



\subsection*{Sensing Hardware Implementation}
In order to forward the analog movements from the foldable structures into quantifiable, digital measurements, we developed a Data Acquisition Unit (DAU) for collecting the movements through capacitive changes.
The DAU consists of two Texas Instruments FDC2214 4-channel 28-bit capacitive-digital converters and an Adafruit Feather
Sense Microcontroller to gather the data streams and forward them through Bluetooth Low Energy.
In total, we are able to collect up to 8-channel capacitive data in parallel through one DAU.
The sampling rate can be flexibly adjusted, whereas we decided to fix it to 30 Hz across our experiments to match the required data resolution and speed of motion.
To gather the capacitive channel data for the machine learning training, we created a JavaScript application to connect with the DAU via Bluetooth Low Energy.
The user interface provides raw signal visualizations to monitor the data collection process and features to pack the acquired data into files.

Throughout the experiment, each channel of the DAU was connected to one of the foldable structure conductive patches through wired pin connections for fabric samples, and double-sided conductive tapes (TESA 60257) for paper-based samples.
Due to the sensitivity and nature of the single-end capacitor, the wired connections movement may affect the signal quality, wherefore we fixed the DAU and the wires to the experiment setup to isolate the foldable structure movement.

Additionally, we set up our experiment with a static single camera (iPhone 12 mini) facing the moving foldable structure to extract the marker positions through image-based tracking.

\subsection*{Data Collection and Preparation Details}
Each session typically consists of 3 to 4 sessions, ranging from 8 to 15 minutes, depending on the complexity of the geometric structure being analyzed.
Motion patterns, comprising arbitrary sequences and combinations of elements shown in \cref{fig:experiment_motions} were applied.

Realistic movements were simulated for both configurations, with intentional avoidance of clear separations between motion elements. The fixed edges of the samples measured 22.5 cm, while the foldable edges (top and bottom) spanned a range from 2cm to 20cm for the Accordion (P,R) and Chevron (P,R) patterns. Whereas V-Fold and Sunray patterns measured around 10 cm with foldable edges spanning a range from 2 cm to 14 cm and 5 cm to 18 cm respectively. 
Each sample, regardless of its placement, underwent recording under these motion conditions for four sessions, each lasting 15 to 20 minutes.

Visual markers on the patches aid in motion tracking and facilitating the deduction of predetermined geometry primitives, which is done in Blender 3D. \cite{blender} 

The FDC2214 utilized in the DAU generates raw capacitive data as frequencies, which was around 13.7 MHz.
As a standard practice, each capacitive channel operates on a slightly different frequency to avoid cross-talking between channels.
To normalize each channel stream, we first subtracted the mean signal of each channel and normalized them afterwards into the range of 0 to 1.
This process shares similarities with a High-Pass filter in order to resolve long-term signal drift and operation frequency while maintaining the signal features of foldable structure movement.

We tracked the Visual markers as 2D patches using Blender VFX motion tracking tool using Fast-Marker preset where we monitor the position, rotation and scaling of a given patch across the frame.

Afterwards, we aligned the capacitive sensor data and the tracked ground truth through their unix timestamps to align the two data sources for the final dataset.
Due to the sampling frequency of 30 Hz, we require millisecond synchronization to align each frame properly.
To resolve possible deviations from unix timestamps due to the different recording devices, we additionally applied unique synchronization pattern movements to finally align the marker tracking movement with the capacitive signal through manual adjustment.

\subsection*{Machine Learning Implementations}

We then trained a deep convolution neural network (CNN), following the CNN regressor architecture proposed in \cite{zhou2023mocapose}, to predict the 3 length primitives from the N capacitive channels (depending on the pattern).

A deep 1D convolutional CNN regressor architecture, as proposed through works like \cite{zhou2023mocapose}, was implemented in PyTorch as shown in \cref{fig:pipeline}c.
The lightweight model has only 27,315 trainable parameters.
The regressor takes 1-second window of normalized capacitive data (30 timesteps $\times$ N channels) and predicts a single timestamp (the middle frame) of the 1-second window. 
The algorithmic latency thus was 0.5 seconds.
The dataset was generated by a sliding window of 30 timestep window size and 1 timestep of window step.

For each sample, an individual model was trained with the first three recordings ($\approx$ 60 minutes) and tested on the last recording ($\approx$ 20 minutes) to avoid overfitting.
The neural network was trained and tested on a workstation with a GPU accelerator (Nvidia RTX 6000 Ada Lovelace) with a batch size of 4096, Adam optimizer with an initial learning rate of 0.01 and applied decay of 0.5 every 20 epochs.
The early stopping metric to terminate the training procedure was set to a patience of 100 epochs.

\subsection*{Availability of Data and Materials}
The datasets generated and/or analyzed during the current study are available from the corresponding author upon reasonable request.

\subsection*{Image Publication Agreement}
Informed consent was obtained from the involved individuals to use their facial image for this publication, with an understanding that it may be disseminated in print, digital, or online formats for the specified purpose.

\bibliography{references}


\section*{Acknowledgements}

The research reported in this paper was supported by the BMBF (German Federal Ministry of Education and Research) in the project VidGenSense (01IW21003).

\section*{Author Contributions Statement}
L.R., D.G., B.Z., B.G. : Methodology, and Writing the Original Draft;
B.G, : Design;
B.Z. : Conceptualization;
D.G. : Electronics;
L.R., D.G. : Data Collection;
L.R. : Ground truth estimation, 3D Simulation;
L.R., D.G., B.Z. : Model Training;
L.R., B.Z. : Digital Art;
P.L.: Funding Acquisition and Supervision;

\section*{Additional Information}
\textbf{Competing interests:} The authors declare no competing financial interests.




\end{document}